\documentclass[12pt]{article}
\usepackage[a4paper, total={7in, 10in}]{geometry}
\usepackage[parfill]{parskip}
\usepackage{physics, tensor, float, subcaption}
\usepackage{graphicx}
\graphicspath{ {Plots/} }
\usepackage{jhep-mod}
\usepackage{bm}
\usepackage{soul}
\usepackage{amssymb,amsmath,amsthm}
\usepackage{mathrsfs}
\usepackage[utf8]{inputenc}
\usepackage{enumerate}
\usepackage{bigints}
\usepackage{xcolor}
\usepackage{appendix}
\usepackage{graphicx}
\usepackage{float}
\usepackage{tikz}
\usepackage{setspace}
\usepackage{cancel}
\usepackage{array}
\usepackage{tabulary}
\usepackage{doi}
\usepackage{comment}
\definecolor{purple}{rgb}{1,0,1}
\definecolor{lime}{HTML}{A6CE39} % needs xcolor
\newcommand{\blue}[1]{{\color{blue} #1}}

\definecolor{lime}{HTML}{A6CE39}
\newcommand{\orcidicon}{%
	\begin{tikzpicture}
	\draw[lime, fill=lime] (0,0) 
		circle [radius=0.16] 
		node[white] {{\fontfamily{qag}\selectfont \tiny ID}};
	\draw[white, fill=white] (-0.0625,0.095) 
		circle [radius=0.007];
	\end{tikzpicture}
	\hspace{-5mm}
}
\newcommand\orcidRudeep{{\href{https://orcid.org/0009-0002-0162-562X}{\orcidicon}}}
\newcommand\orcidMatt{{\href{https://orcid.org/0000-0003-1088-6485}{\orcidicon}}}
\newcommand{\be}{\begin{equation}}
\newcommand{\ee}{\end{equation}}

\begin{document}
%========================================================

%========================================================
%========================================================

\title{\vspace{-75pt}\huge
Black holes, white holes, \\
and near-horizon physics
}

%========================================================
%========================================================

%========================================================
\author{
\Large
Rudeep Gaur\!\orcidRudeep {\sf  and} Matt Visser\!\orcidMatt}
%========================================================
%========================================================
%========================================================
%========================================================
\affiliation{School of Mathematics and Statistics, Victoria University of Wellington, 
\\
\null\qquad PO Box 600, Wellington 6140, New Zealand.}
%========================================================
%========================================================
\emailAdd{rudeep.gaur@sms.vuw.ac.nz}
\emailAdd{matt.visser@sms.vuw.ac.nz}
%========================================================
%========================================================

\abstract{
\vspace{1em}

Black and white holes play remarkably contrasting roles in general relativity \emph{versus}\\ observational astrophysics. 
While there is observational evidence for the existence of compact objects that are ``cold, dark, and heavy'',  which thereby  are natural candidates for black holes, the theoretically  viable time-reversed variants --- the ``white holes'' --- have nowhere near the same level of observational support. Herein we shall explore the theoretical possibility that the connection between black and white holes is much more intimate than commonly appreciated. We shall first construct ``horizon penetrating'' coordinate systems that differ from the standard curvature coordinates only in a small near-horizon region, thereby emphasizing that ultimately the distinction between black and white horizons depends only on near-horizon physics.
We shall then construct an explicit model for a ``black-to-white transition'' where all of the nontrivial physics is confined to a compact region of spacetime --- a finite-duration finite-thickness, (in principle arbitrarily small), region straddling the na{\"i}ve horizon. Moreover we shall show that it is possible to arrange the ``black-to-white transition'' to have zero action --- so that it will not be subject to destructive interference in the Feynman path integral. 
This then raises the very intriguing possibility that astrophysical black holes might be interpretable in terms of a quantum superposition of black and white horizons --- a ``gray'' horizon.

\bigskip
\noindent
{\sc Date:} 4 April 2024; 10 May 2024; \LaTeX-ed \today

\bigskip
\noindent
{\sc arXiv:}  2304.10692 [gr-qc]

\bigskip
\noindent
{\sc Published as:} JHEP {\bf 2024} (2024) 172
%Volume 2024, article number 172, (2024)

\bigskip
\noindent
{\sc doi:} 10.1007/JHEP05(2024)172

black-to-white transitions; quantum superpositions.

}

%========================================================
\maketitle
%========================================================
\def\tr{{\mathrm{tr}}}
\def\diag{{\mathrm{diag}}}
\def\cof{{\mathrm{cof}}}
\def\pdet{{\mathrm{pdet}}}
\parindent0pt
\parskip7pt
%=====================================================
%=====================================================
\section{Introduction}
%=====================================================

Classical black holes are objects that, from a theoretical perspective, are very well understood within the standard framework of the theory of general relativity~\cite{Weinberg,MTW,Adler-Bazin-Schiffer,Wald,D'Inverno,Hartle,Carroll,Hobson}. \\
Likewise, the observational~\cite{EHT-1,EHT-4,EHT-5,EHT-SgA*-1,EHT-SgA*-6} and phenomenological~\cite{Psaltis:2014,Broderick:2013,Cardoso:2016,Carballo-Rubio:2018a,Carballo-Rubio:2018b} situations are both increasingly well understood.
The (mathematical) event horizon, or the physically more relevant long-lived apparent horizon~\cite{weather,observability}, is often dubbed ``the point of no return" and is not really a problematic issue under suitable coordinate choices. However, one certainly  finds that the central singularity still causes many conceptual problems with our understanding of physics. One of the most prominent problems being the destruction of information as it approaches the singularity. Some of the theories that are put forward to resolve the information paradox are soft hairs that evaporate to null infinity, and \textit{white holes}. While in this paper we will not delve into the information paradox itself, it is important to understand some of the motivation behind white holes. 
A representative selection of references includes~\cite{Ashtekar:2018,Macher:2009,Barrau:2014,Eardley:1974,Barcelo:2015,Rovelli:2018,Wald:1980,Rovelli:2018b,McClure:2008,Zaslavskii:2017,Barrau:2021,Nikitin:2018,Barcelo:2015b,Hsu:2010,Lake:2011,Kedem:2020,Volovik:2005,Gomez:2002,Retter:2011,Garay:2017,Bishop:2009,Jannes:2012}.

White holes, as the name may suggest, are hypothesised to be the opposite of black holes; a ``time reversed'' black hole. Matter is radiated from the horizon instead of being absorbed thereby. There are many theories as to how white holes might form from black holes, most of which involve some sort of quantum mechanical effect. A representative selection of references includes~\cite{Hajicek:2001, Hajicek:2000, Haggard:2014,Bianchi:2018,Olmedo:2017,Barcelo:2014,DeLorenzo:2015,Bodendorfer:2019,Barcelo:2014b,Barcelo:2015c,BenAchour:2020,Barcelo:2016,Christodoulou:2018,Martin-Dussaud:2019,Maciel:2015, Brahma:2018,Bardeen:2018,Haggard:2015b,Rignon-Bret:2021,Han:2023, Hong:2022,Bardeen:2020,Jalalzadeh:2022,Starobinsky:1975,Volovik:2021}.

One specific example of this phenomenon can be found in reference~\cite{Hajicek:2001}, where the authors discuss ``gray'' horizons --- as hypothetical quantum superpositions of black and white horizons.
Another example can be found in reference~\cite{Bianchi:2018} where those authors hypothesise that black holes quantum tunnel into white holes once a black hole evaporates down to the Planck mass. Other theories, such as those proposed in references \cite{Haggard:2014,Barcelo:2015c}, involve modifying large wedges of the spacetime (typically all the way down to the central singularity) in order to have a black hole ``bounce" to a white hole. 

Herein we will propose simple and explicit fully \textit{classical} models for a white hole, and in particular for a black-to-white hole transition. 
\begin{itemize}
\item 
Firstly, starting from the standard (Hilbert) form of the Schwarzschild metric in curvature coordinates, we shall introduce a simple coordinate change, through a function depending solely on the radial coordinate, $r$. Specific choices of this function will result in a static black hole and white hole in horizon-penetrating coordinates --- such as Painl\'eve--Gullstrand, Kerr--Schild, and Eddington--Finkelstein coordinates. 
\item
Secondly,  we shall localize the required coordinate change to a compact near-horizon radial region, showing that both black and white holes can be cast into the standard manifestly static form outside of some compact radial region. Thus a clean distinction can be made between ``black'' and ``white'' horizons with minimal modifications to the standard (Hilbert) form of the Schwarzschild metric. 
\item 
Thirdly, we introduce a function of time to create a non-vacuum spacetime, one that is no longer static, and which describes a black to white hole ``bounce''; with the ``bounce'' being confined to a compact (arbitrarily small) region of spacetime. Furthermore, an analysis of the action in the transition region will be conducted,  the radial null curves will be investigated, and various energy conditions will be checked. 
Finally, we shall connect the discussion to quantum physics by applying the  Feynman functional integral approach. 
\end{itemize}

\enlargethispage{35pt}
Our approach will only require fine tuning of the Schwarzschild spacetime in a compact radial region \textit{near the horizon}. Therefore, the entire spacetime outside of a small neighbourhood of $r = 2m$ will be that of the standard (Hilbert) form of Schwarzschild spacetime. This is achieved by the use of smooth bump functions that will not create discontinuities in the metric; and therefore the Christoffel symbols  will not be discontinuous, and the Riemann tensor will not contain delta-function contributions.

%------------------------------------------------------
\section{Static black and white horizons: Global analysis}
%-------------------------------------------------------

Firstly, we will introduce a particularly simple model for (static) black and white horizons, by performing some absolutely minimal modifications of standard textbook results. We begin with the Schwarzschild spacetime (in the usual Hilbert/curvature coordinates):
\begin{equation}
ds^2 = -\left(1-{2m\over r}\right) dt^2 + {dr^2\over1-2m/r} + r^2 d\Omega^2.
\end{equation}
Using the following coordinate transformation,
\begin{equation}
t \to t + F(r); \qquad dt \to dt + f(r) dr,
\end{equation}
results in the line element
\begin{equation}
ds^2 = -\left(1-{2m\over r}\right) (dt+f(r)dr)^2 + {dr^2\over1-2m/r} + r^2 d\Omega^2.
\end{equation}
Expanding, this implies
\begin{equation}
\begin{split}
ds^2 =& -\left(1-{2m\over r}\right) dt^2 - 2 (1-2m/r) f(r) dr dt \\
&+ \left[{1\over1-2m/r} -(1-2m/r) f(r)^2 \right] dr^2 + r^2 d\Omega^2.
\end{split}
\end{equation}
It is important to note that this line element is still Ricci flat, and so is merely the Schwarzschild geometry in disguise,  for \emph{arbitrary} $f(r)$.

Without any loss of generality, one may choose:
\begin{equation}
f(r) = {h(r)\over1-2m/r}.
\end{equation}

This then results in the line element
\begin{equation}
ds^2 = -\left(1-{2m\over r}\right)  dt^2 - 2 h(r) dr dt + \left[{1- h(r)^2 \over1-2m/r} \right] dr^2 + r^2 d\Omega^2.
\end{equation}
\emph{All} of these line elements, for arbitrary $h(r)$, are just (coordinate) variants of the standard Schwarzschild spacetime --- they are all Ricci-flat for \emph{arbitrary} $h(r)$. 
For specific choices for the function $h(r)$ we obtain some particularly well known coordinate variants of the Schwarzschild spacetime.

%=========================================
\subsection{Painl\'eve--Gullstrand coordinates}
%==========================================

Set $h(r)\to \pm \sqrt{2m/r}$, then
\begin{equation}
ds^2 = -\left(1-{2m\over r}\right)  dt^2 \mp 2 \sqrt{2m/r} \; dr dt +  dr^2 + r^2 d\Omega^2.
\end{equation}
Examining the radial null condition, $ -dt^2 +\left(dr\mp  \sqrt{2m/r} \;dt\right)^2 =0$, 
we see that in this coordinate system the radial null curves are
\begin{equation}
{dr\over dt} = \pm 1 \pm  \sqrt{2m/r},
\end{equation}
where the signs are to be chosen independently. 
\begin{itemize}
\item 
Thence for a black hole we choose
\begin{equation}
{dr\over dt} = \pm 1 -  \sqrt{2m/r},
\end{equation}
with ${dr\over dt} \in \{ 0,-2\}$ at horizon crossing ($r=2m)$. 
\item
In contrast for a white hole we choose
\begin{equation}
{dr\over dt} = \pm 1 +  \sqrt{2m/r},
\end{equation}
with ${dr\over dt} \in \{ +2,0\}$ at horizon crossing ($r=2m)$.
\end{itemize}

%======================================
\subsection{Kerr--Schild coordinates}
%======================================
Set $h(r)\to \pm {2m/r}$, then
\begin{equation}
ds^2 = - dt^2 + dr^2 + r^2 d\Omega^2 + {2m\over r} (dt \pm dr)^2.
\end{equation}
Examining the radial null condition, $ -dt^2 +dr^2 + (2m/r) (dt\pm dr)^2 =0$,  
in this coordinate system we find the radial null curves are either
\begin{equation}
{dr\over dt} = \mp1, \quad\text{or}\quad {dr\over dt} = \pm1 \mp{4m\over r+2m},
\end{equation}
where the signs are to be chosen in a correlated manner. 
\begin{itemize}
\enlargethispage{20pt}
\item 
Thence for a black hole we choose either
\begin{equation}
{dr\over dt} = -1\quad\hbox{(ingoing)}, 
\qquad\text{or}\qquad 
{dr\over dt} = 1   -{4m\over r+2m} \quad\hbox{(``outgoing'')}, 
\end{equation}
with ${dr\over dt} \in \{ -1,0\}$ at horizon crossing ($r=2m)$. 
\item
In contrast for a white hole we choose either
\begin{equation}
{dr\over dt} = 1  \quad\hbox{(outgoing)}; 
\qquad\text{or}\qquad 
{dr\over dt} =  -1  +{4m\over r+2m} \quad\hbox{(``ingoing'')},
\end{equation}
with ${dr\over dt} \in \{1,0\}$ at horizon crossing ($r=2m)$.
\end{itemize}

%=============================================
\subsection{Eddington--Finkelstein null coordinates}
%=============================================
Set $h(r) = \pm 1$, then 
\begin{equation}
ds^2 = -(1-2m/r) dt^2 \mp  2 dr dt + r^2 d\Omega^2\,.
\end{equation}
Depending on the choice of sign, $\pm$, one usually relabels $t\to u$ or $t\to v$.
\begin{itemize}
\item 
The ingoing Eddington--Finkelstein coordinates are typically given as
\begin{equation}
ds^2 = -(1-2m/r) dv^2 +  2 dv dr + r^2 d\Omega^2\,.
\end{equation}
Examining the radial null condition, $[ -(1-2m/r) dv +  2 dr ] dv =0$, 
and noting that this quantity must be negative for timelike curves, 
we find the radial null curves are 
\begin{equation}
{dr\over dv} = - \infty; \qquad {dr\over dv} = {1-2m/r\over 2}.
\end{equation}
Thence the ingoing Eddington--Finkelstein coordinates represent a black hole
with ${dr\over dv} \in \{ -\infty,0\}$ at horizon crossing ($r=2m)$. 

\item
The outgoing Eddington--Finkelstein coordinates are typically given as
\begin{equation}
ds^2 = -(1-2m/r) du^2 -  2 du dr + r^2 d\Omega^2\,.
\end{equation}
Examining the radial null condition, $ [-(1-2m/r) du-  2 dr ]du =0$, 
and noting that this quantity must be negative for timelike curves, 
we find the radial null curves are 
\begin{equation}
{dr\over du} = + \infty; \qquad {dr\over du} = - {1-2m/r\over 2}.
\end{equation}
Thence the outgoing Eddington--Finkelstein coordinates represent a white hole
with ${dr\over du} \in \{ +\infty,0\}$ at horizon crossing ($r=2m)$. 

\end{itemize}

%===========================================================
\subsection{Generic horizon-penetrating coordinates} 
%===========================================================

\enlargethispage{30pt}
From the above we see that all three of these coordinate systems, Painl\'eve--Gullstrand,  Kerr--Schild, and Eddington--Finkelstein provide three specific \emph{examples} of horizon-penetrating coordinates. In each case, depending on whether one is in a black hole or a white hole configuration, one of the radial null geodesics remains frozen on the horizon --- i.e., the coordinate velocity is zero --- while the other crosses the horizon with a non-zero coordinate velocity. 

Of course there are infinitely many other horizon-penetrating coordinates~\cite{Sarbach:2001,Campanelli:2000, Bhattacharyya:2020,Cherubini:2018,Boonserm:2017,Cherubini:2023}, some of which we explore below, these three \emph{examples} are just three of the most obvious ones. 
We can make the required coordinate transformations fully explicit by noting
\begin{equation}
F(r) = \int f(r)\; dr = \int {h(r)\over1-2m/r} \; dr.
\end{equation}

Then,  for these three specific examples, we see
\begin{equation}
F_{PG}(r) =  \pm \int {\sqrt{2m/r} \over1-2m/r} \; dr = \pm2\sqrt{2mr} \pm 2m \ln\left(1-\sqrt{2m/r}\over1+\sqrt{2m/r}\right);
\end{equation}
\begin{equation}
F_{KS}(r) =  \pm \int {2m/r \over1-2m/r} \; dr = \pm2m\ln(r-2m);
\end{equation}
\begin{equation}
F_{EF}(r) =  \pm \int {1\over1-2m/r} \; dr = \pm r \pm 2m\ln(r-2m).
\end{equation}
These three functions all share the feature of being somewhat unpleasantly behaved near spatial infinity. Specifically, for these three coordinate systems one has (perhaps unexpectedly)  to make unboundedly large alterations to the time coordinate near spatial infinity, where the gravitational field is weak.  Such behaviour, while not fatal, is perhaps somewhat annoying --- we shall first seek to ameliorate it by keeping the function $h(r)$ finite and localized to a compact region thereby keeping the function $f(r)$ integrable, and the function $F(r)$ bounded.

%------------------------------------------------------
\section{Static black and white horizons: Local analysis}
%-------------------------------------------------------

We now let $h(r)$ be a bump function. 
At the horizon, pick $h(2m)=\pm 1$, with $h(r)$ being some finite smooth function of compact support.
Then we have a version of the Schwarzschild line element presented with localized version of horizon penetrating coordinates.
At $r=2m$ there is either a black or white horizon depending on the \emph{sign} of $h(2m)$.
This line element goes to the standard Hilbert form of Schwarzschild at some finite $r$, (both large and small $r$).
That is: $support\{h(r)\} \subseteq [r_<, r_>]$, with $2m \in (r_<, r_>)$.
This is still a Ricci-flat coordinate transformed version of Schwarzschild:
\begin{equation}
ds^2 = -\left(1-{2m\over r}\right)  dt^2 - 2 h(r) dr dt + \left[{1- h(r)^2 \over1-2m/r} \right] dr^2 + r^2 d\Omega^2.
\end{equation}\enlargethispage{30pt}
Note specifically that to get horizon-penetrating coordinates, (and so obtain either an explicitly black or explicitly white horizon), you only need to adjust the coordinates in the immediate vicinity of the horizon. ``Global'' changes to the coordinates are by no means necessary.

\clearpage
We check the ingoing/outgoing null curves to verify that the coordinates are in fact horizon penetrating. 
We have
\begin{equation}
 -\left(1-{2m\over r}\right)  dt^2 - 2 h(r) dr dt + \left[{1- h(r)^2 \over1-2m/r} \right] dr^2 = 0.
\end{equation}
Thence
\begin{equation}
 -\left(1-{2m\over r}\right)^2   - 2 \left(1-{2m\over r}\right) h(r) \dot r+ \left[{1- h(r)^2} \right] \dot r^2 = 0.
\end{equation}

This is an easily solved quadratic for $\dot r$, leading to 
\begin{equation}
\dot r = \mp {1-2m/r\over 1\pm h(r)}.
\end{equation}
Depending on the (implicit) sign choice hiding in $h(2m)=\pm 1$, and the explicit sign choice $\pm$ multiplying $h(r)$, one of these null curves will be trapped at the horizon, (with $\dot r_H=0$), while the other null curve crosses the horizon with a coordinate speed that is formally $0/0$, and so must be determined by using the l'H\^opital rule:
\begin{equation}
\dot r_H = \pm {1\over 2m \; h'(2m)}.
\end{equation}
Therefore, we find these are generically horizon-penetrating coordinates. (At least  \emph{one} of the radial null curves has nonzero coordinate velocity at horizon crossing). 
The net amount by which we have to adjust the time coordinate to achieve this localized horizon-penetrating behaviour is
\begin{equation}
\label{E:shift}
\Delta F = F(\infty)-F(0) = 
F(r_>)-F(r_<) = \int_{r_<}^{r_>} {h(r) \over1-2m/r} \; dr\;
= \int_{\blue{r_<}}^{\blue{r_>}} {r \,h(r) \over r-2m} \; dr.
\end{equation}
The na{\"i}ve singularity at the horizon $r=2m$ is an integrable singularity, so the net shift in the  time coordinate is finite. 

%=============================================
\section{Black-to-white bounce: Compact transition region}
%==============================================

We now wish to move away from consideration of static black and white holes, and explore a classical model of a black-to-white hole transition. To do so, we make the following change:
\begin{equation}
h(r) \to s(t)\; h(r).
\end{equation}
This is no longer \textit{just} a coordinate transformation. The spacetime is no longer Ricci-flat. 
Specifically, we consider the metric
\begin{equation}
ds^2 = -(1-2m/r) dt^2 - 2 s(t) h(r) dr dt + \left[{1- s(t)^2 h(r)^2 \over1-2m/r} \right] dr^2 + r^2 d\Omega^2.
\end{equation}
We again take $h(2m)=\pm1$, and take $h(r)$ to be of compact support, that is:  $support\{h(r)\} \subseteq [r_<, r_>]$.
Furthermore we shall also assume that $1-s(t)^2$ is of compact support with $s(t)\to \pm1$ at large $|t|$. In fact we shall take $s(+\infty)=\pm 1$ and $s(-\infty)=\mp 1$, since we want to enforce a sign flip in $s(t)$ to enforce a black-to-white transition.
That is, $support\{1-s(t)^2\} \subseteq [t_<, t_>]$. 
This in turn implies $support\{\dot s(t) \} \subseteq [t_<, t_>]$. 
We again emphasize: this geometry is not Ricci flat --- it is no longer \textit{just} a coordinate transformation.\footnote{Somewhat similar constructions can be found in references~\cite{Barcelo:2014b,Barcelo:2016}.} 

%\clearpage

%-----------------------------------------------
\subsection{Einstein tensor}
%===========================
Since the spacetime is not just a coordinate transformation of the Schwarzschild metric, the Einstein tensor and Ricci tensor will now be non-zero.
We calculate the Einstein tensor, ({\sf Maple}), its non-zero radial-temporal components are \\
\begin{equation}
G_{tt}=0;
\qquad\qquad
G_{rr} = - {2\dot s(t) h(r)\over r(1-2m/r)};
\end{equation}
while the orthonormal angular components are
\begin{equation}
G_{\hat\theta\hat\theta} = G_{\hat\phi\hat\phi} =  {d^2[s^2(t)]/dt^2 \; h(r)^2\over 2(1-2m/r)} {+} h'(r) \dot s(t) - {(1-m/r) h(r) \dot s(t) \over r (1-2m/r)}.
\end{equation}
The Ricci scalar is
\begin{equation}
R = {-}  {d^2[s^2(t)]/dt^2 \; h(r)^2\over (1-2m/r)} {-} 2 h'(r) \dot s(t) {+} {2(2-3m/r) h(r) \dot s(t) \over r (1-2m/r)}.
\end{equation}
Note the Einstein tensor is of compact support --- it is only nonzero where \emph{both} $h(r)$ \emph{and the derivatives} $\{\dot s(t),\ddot s(t)\}$ are non-zero.

Note that both the metric determinant, $g=-r^4\sin^2\theta$, and the volume element, $\sqrt{-g}=r^2\sin\theta$, are independent of both $h(r)$ and $s(t)$.

%===============================================
\subsection{Finite action for the bounce}
%===============================================
The contribution to the action from the transition region is finite. First we note
\begin{equation}
S = \int \sqrt{-g} \; R \; d^4x =  \int \sqrt{-g} \; R \; d^4x = 4\pi \int r^2 \; R \; dt dr.
\end{equation}
But the $t$ integration yields 
\begin{equation}
\int_{-\infty}^{+\infty} \left(d^2[s^2(t)]\over dt^2\right)  \; dt = \left[d[s^2(t)]\over dt\right]_{-\infty}^{+\infty}= 0-0= 0,
\end{equation}
and 
\begin{equation}
\int_{-\infty}^{+\infty} \left(ds(t)\over dt\right)  \; dt = \left[s(t)\right]_{-\infty}^{+\infty}=  \pm 1 - (\mp 1) = \pm 2.
\end{equation}

Therefore 
\begin{equation}
S = \pm
4\pi \int r^2 \left[ 4 h'(r) +  {4(2-3m/r) h(r)  \over r (1-2m/r)}\right]  dr.
\end{equation}

Now integrate by parts in the radial coordinate
\begin{equation}
\int_{-\infty}^{+\infty} r^2 h'(r) dr = \left[ r^2 h(r) \right]_{-\infty}^{+\infty} - \int_{-\infty}^{+\infty} 2 r h(r) dr = - \int_{-\infty}^{+\infty} 2 r h(r) dr
\end{equation}

Therefore 

\begin{equation}
S = 
\pm 4\pi \int r \left[ -8 h(r) +  {4(2-3m/r) h(r)  \over (1-2m/r)}\right]  dr.
\end{equation}
After some algebra, this is explicitly:
\begin{comment}
That is 
\begin{equation}
S = \pm 16\pi \int r \left[ {m/r  \over (1-2m/r)}\right]  h(r) dr.
\end{equation}
That is 
\begin{equation}
S = \pm 16\pi m \int { h(r) \over (1-2m/r)}\;   dr.
\end{equation}
Even more explicitly
\end{comment}
\begin{equation}
\label{E:S}
S = \pm 16\pi m \int_{r_<}^{r_>} { h(r) \over (1-2m/r)}\;   dr
= \pm 16\pi m \int_{r_<}^{r_>} { r\,h(r) \over (r-2m)}\;   dr.
\end{equation}
(The na{\"i}ve singularity at $r=2m$ is again an integrable singularity.)
While the interpolating spacetime geometry is now dynamic---not static---the total action can be written in terms of the time-shift (\ref{E:shift}) at late and early times,  (when the geometry is static),  as
\begin{equation}
S = \pm 16\pi m \;  \Delta F.
\end{equation}\enlargethispage{40pt}
The reason the finiteness of the action  is important is that finite-action configurations can easily contribute non-destructively to the Feynman path-integral.
(The contributions of infinite action configurations tend to `wash out' due to destructive interference.)

%===============================================
\subsection{Zero action for the bounce}
%===============================================

Perhaps unexpectedly, by making a suitable (symmetric) choice for $h(r)$ we can even drive the action of our black-to-white bounce to zero, not just keeping it finite.

For example:  Take $ r_{>}= 2m + \Delta$, and $ r_{<}= 2m - \Delta$, and subsequently choose $h(r) = \pm (2m/r) B(|r-2m|)$; where $B(x)$ is a bump function with $B(0)=1$ and $B(\Delta) = 0$; in this static case this leads to coordinates that are locally Kerr--Schild in the immediate vicinity of the horizon.

Then for the action of the black-to-white bounce, after integrating out the time dependence, from (\ref{E:S}) we have:
\begin{equation}
S = \pm 16\pi m \int_{r_<}^{r_>} { h(r) \over (1-2m/r)}\;   dr = \pm 16\pi m \int_{2m-\Delta}^{2m+\Delta} { (2m/r) B(|r-2m|) \over (1-2m/r)}\;   dr
\end{equation}
\begin{equation}
= \pm 32\pi m^2 \int_{2m-\Delta}^{2m+\Delta} { B(|r-2m|) \over (r-2m)}\;   dr
= \pm 32\pi m^2 \int_{-\Delta}^{+\Delta} { B(|z|) \over z}\;   dz.
\end{equation}
Here we have defined $z=r-2m$, 
This integral obviously vanishes by symmetry, but for clarity, being  careful with the integrable singularity
\begin{equation}
S \propto  \lim_{\epsilon\to0} \left( \int^{-\epsilon}_{-\Delta} { B(|z|) \over z}\;   dz  + \int^{\Delta}_{\epsilon} { B(|z|) \over z}\;   dz \right).
\end{equation}
Thence
\begin{equation}
S \propto  \lim_{\epsilon\to0} \left( \int_{\epsilon}^{\Delta} { B(|z|) \over z}\;   dz  - \int^{\Delta}_{\epsilon} { B(|z|) \over z}\;   dz \right) = 0.
\end{equation}
We may therefore conclude that one can even construct a \emph{zero-action} compact support Lorentzian ``bounce'' that converts black holes to white holes (and \emph{vice versa}).

%===============================================
\subsection{Radial null curves}
%===============================================

The radial null curves in this time dependent geometry are specified by
\begin{equation}
-(1-2m/r) dt^2 - 2 s(t) h(r) dr dt + \left[{1- s(t)^2 h(r)^2 \over1-2m/r} \right] dr^2 =0.
\end{equation}
That is 
\begin{equation}
-(1-2m/r)^2  - 2 s(t) h(r) (1-2m/r) \dot r + [1- s(t)^2 h(r)^2] \dot r^2 =0.
\end{equation}
This is a simple quadratic for $\dot r$,  implying
\begin{equation}
{dr\over dt} = \pm  {(1-2m/r)\over   [1\mp s(t) h(r)]}.
\end{equation}
Unfortunately this ODE is not separable, and is not easy to solve.

\enlargethispage{40pt}
The radial null curves are of the form 
\begin{equation}
k^a \propto (1, \dot r; 0, 0) =\left(1, \pm  {(1-2m/r)\over   [1\mp s(t) h(r)]}; 0,0\right).
\end{equation}
In regions where $s(t)^2=1$, and using the fact that we always impose $h(2m)=1$, one or the other of these radial null curves  will be horizon penetrating. (In particular at early and late times, where $|s(t)|=1$, one or the other of the null curves will penetrate the na{\"i}ve horizon.)

During the bounce we can for simplicity assert $|s(t)|<1$, and in fact $s(t)$ must, by construction, pass through zero.
We can also for simplicity assert $|h(r)|\leq 1$, with equality only at the na{\"i}ve horizon $r=2m$. Under these conditions the denominator $1\mp s(t) h(r)$ is always nonzero and both incoming and outgoing null rays will be (temporarily) trapped at the na{\"i}ve horizon, both with $\dot r_H=0$ --- at least until the end of the bounce --- when, as per our analysis above, one or the other null curve can cross $r=2m$ with nonzero coordinate velocity. 

%==================================
\subsection{Energy conditions}
%==================================

While it is by now clear that the classical point-wise energy conditions of general relativity are not truly fundamental~\cite{twilight,Curiel:2014,Kontou:2020}, (since they are all violated to one extent or another by quantum effects~\cite{gvp1,gvp2,gvp3,gvp4,gvp5}), they are nevertheless extremely good diagnostics for detecting ``unusual physics'' that merits a very careful examination~\cite{Hochberg:1998,Lobo:2020,Santiago:2021,Visser:2021}. 
The status of integrated energy conditions~\cite{Wald:1991,Flanagan:1996,scale-anomalies} and quantum inequalities  is much more subtle~\cite{Ford:1994}. 
In the current context it is most useful to focus on the null energy condition (NEC) and trace energy condition (TEC).

\paragraph{NEC:}
The condition for the null energy condition (NEC) to hold is $G_{ab} \;k^a \,k^b \geq 0$. 
The quantity $G_{ab} \;k^a \,k^b$  can be easily calculated for radial null curves, and in this case is:
\begin{equation}
G_{ab}\; k^a \,k^b \propto G_{rr}  \left({(1-2m/r)^2\over   [1\mp s(t) h(r)]^2 }\right) = - {2\dot s(t) h(r) (1-2m/r) \over  [1\mp s(t) h(r)]^2 }.
\end{equation}
Since the denominator is non-negative we see
\begin{equation}
G_{ab}\; k^a\, k^b \propto  - {\dot s(t)\, h(r)\, (1-2m/r) }.
\end{equation}
Regardless of the sign of $\dot s(t)$, or the sign of $h(2m)$,  the product $\dot s(t) \; (1-2m/r) $ will certainly flip sign as one crosses the na{\"i}ve horizon at $r=2m$. Therefore, the NEC  is definitely violated in parts of the black-to-white transition region. Furthermore, this automatically implies that the WEC, SEC, and DEC are also violated in parts of the black-to-white transition region.

\paragraph{TEC:} The trace energy condition (TEC) is important mainly for historical reasons~\cite{twilight}, though there is currently some resurgence of interest in this long-abandoned energy condition. (The TEC is useful for ordinary laboratory matter, but is already known to be violated by the equation of state for the material in the deep core of neutron stars, and in fact for any ``stiff'' system where $w\equiv p/\rho$ exceeds ${1/\sqrt{3}}$.)

\enlargethispage{40pt}
The TEC asserts
\begin{equation}
g_{ab} \, T^{ab} = -(\rho-3p) \leq 0.
\end{equation}
For the Einstein tensor this becomes $g_{ab}\, G^{ab}\leq0$, and for the Ricci scalar $R \geq  0$. But this would imply a positive semidefinite action, and we know that the black-to-white transition region is non-vacuum and can be chosen to have zero action. Thence there must certainly be regions in the compact black-to-white transition region where the TEC is violated. 

\paragraph{ANEC:} Analyzing the averaged null energy condition (ANEC) would require one to trace the null geodesics through the bounce region, and to unambiguously identify a suitable null affine parameter. Unfortunately, this is one of those situations where (despite recent progress~\cite{null-affine}) these issues are still in the ``too hard'' basket.

\bigskip
Overall, we see that key point-wise energy conditions are definitely violated by the black-to-white bounce. This is an invitation to think carefully about the underlying physics. 

%===============================================
\section{Quantum implications}
%===============================================
 
Despite considerable efforts, we do not as yet have a fully acceptable and widely agreed upon theory of quantum gravity. On the other hand, there are plausible and tolerably well accepted partial models --- such as approximations based on semi-classical gravity (and quantized linearized weak-field gravity for that matter). One issue on which there is widespread agreement is the use of the Feynman functional integral formalism in the semi-classical regime.

%\clearpage
One of the key features of the Feynman functional integral formalism is that quantum amplitudes are dominated by classical configurations (plus fluctuations). In the current context, the fact that we have found zero-action black-to-white bounces, combined with the fact that the usual classical vacuum (Schwarzschild) is also zero-action, 
implies that these configurations reinforce constructively. If the black-to-white bounces are to be quantum mechanically suppressed, such suppression will have to come from the quantum fluctuations, not from the leading order term.  

%\clearpage
This situation is somewhat reminiscent of the role played by instanton contributions to the QCD vacuum~\cite{tHooft:1976,Callan:1976,Peccei:1977,Wilczek:1977}. 
There are significant differences, zero-action \emph{versus} finite action, Lorentzian signature \emph{versus} Euclidean signature ---  but crucial key features are similar. 
Indeed, the existence of localized zero-action configurations is not all that unusual, also occurring in flat Minkowski space classical field theories~\cite{zero-action}, though their implications have not been particularly well studied.

This \emph{suggests} the possibility that astrophysical black holes, (the ``cold, dark, and heavy'' objects detected by astronomers), might be in a quantum superposition of black hole and white hole states. For somewhat similar suggestions, differing in detail, see also~\cite{Haggard:2014,Bianchi:2018,Olmedo:2017,Barcelo:2014,DeLorenzo:2015,Bodendorfer:2019,Barcelo:2014b,Barcelo:2015c,BenAchour:2020,Barcelo:2016,Christodoulou:2018,Martin-Dussaud:2019,Maciel:2015, Brahma:2018,Bardeen:2018,Haggard:2015b,Rignon-Bret:2021,Han:2023, Hong:2022,Bardeen:2020,Jalalzadeh:2022,Starobinsky:1975,Volovik:2021}.
Finally one could \emph{speculate} that this is evidence in favour of quantum physics becoming dominant in near-horizon physics --- this was for many decades (pre-2000 CE) a minority opinion within the general relativity community, as there was a broad but not universal consensus that quantum physics should only come into play in the deep core where curvature reaches Planck scale values. More recently (post-2000 CE) the situation is more nuanced.

One of  the main counterweights to that prior (pre-2000 CE) consensus opinion 
is the ``gravastar''  model~\cite{Mazur:2004,Mazur:2001,Mazur:2001b,
Visser:2003,Cattoen:2005,Chirenti:2007,Carter:2005,Chirenti:2008,
Martin-Moruno:2011,Lobo:2015,Lobo:2012, Carballo-Rubio:2022}, 
where quantum physics kicks in at/near the would-be horizon. 
Similarly for the ``fuzzball'' model, stringy physics~\cite{Mathur:2005, Mathur:2008, Mathur:2009, Skenderis:2008, Raju:2018, Guo:2017} kicks in at/near the would-be horizon.
Furthermore, for the ``firewall'' proposal~\cite{Almheiri:2012, Almheiri:2013,Susskind:2012a,Susskind:2012b, VanRaamsdonk:2013, Page:2013, Saravani:2012, Banks:2012, Chen:2015,  Larjo:2012, Mathur:2013-firewall} something again happens  at/near the would-be horizon.
While these proposals typically severely impact on the spacetime geometry throughout the entire interior region, the novel construction we are dealing with in the current article affects only the near-horizon  spacetime geometry.

%===============================================
\section{Conclusion}
%===============================================
 
Our objective in the above was to investigate if a simple and compelling classical model of a black-to-white hole transition could be found. We began by performing a simple coordinate transformation of the standard Schwarzschild metric by modifying the radial coordinate. This resulted in the line element 
\begin{equation}
    ds^2 = -\left(1-{2m\over r}\right)  dt^2 - 2 h(r) dr dt + \left[{1- h(r)^2 \over1-2m/r} \right] dr^2 + r^2 d\Omega^2\,.
\end{equation}
For specific choices of $h(r)$ this returns the Schwarzschild spacetime in other well known coordinates, such as the Painlev\'e--Gullstand, Kerr--Schild, and Eddington--Finkelstein coordinates. By imposing the restriction $h(2m) = \pm1$ we showed that this line element can model a classical black or white hole where one or the other of the null curves are horizon penetrating with nonzero coordinate velocity
\begin{equation}
    \dot r_H =    \pm {1\over 2m \; h'(2m)}\,.
\end{equation}
By choosing $h(r)$ to be of compact support, we demonstrated that we could confine the non-trivial aspects of black and white horizons to a compact radial region straddling the na{\"i}ve horizon $r=2m$

\enlargethispage{20pt}
By introducing a time-dependent function, $s(t)$, we then produced a simple classical model for a black-hole-to-white-hole transition. This spacetime, however, was no longer \textit{just} a coordinate transformation of Schwarzschild spacetime. 

The introduction of $s(t)$ led to the following line element
\begin{equation}
    ds^2 = -(1-2m/r) dt^2 - 2 s(t) h(r) dr dt + \left[{1- s(t)^2 h(r)^2 \over1-2m/r} \right] dr^2 + r^2 d\Omega^2\,.
\end{equation}
The non-static spacetime in these coordinates was found (at early and late times) to have horizon penetrating null curves with coordinate velocity
\begin{equation}
    \dot r_H = \pm {1\over 2m \;h'(2m)}\,.
\end{equation}
During the bounce itself the behaviour of the null curves is much trickier.

We further showed that the action in the transition region was \textit{finite},
\begin{equation}
    S = 16\pi m \int_{r_<}^{r_>} { h(r) \over (1-2m/r)}\;   dr\,.
\end{equation}
More importantly though, this action can be arranged to be zero by carefully choosing $h(r)$. This proves to be a significant result as this action could then be added to the Feynman path integral and have no impact on any quantum amplitudes. 

For tractability and ease of exposition the current analysis has focussed on the Schwarzschild spacetime, though there is no real difficulty (apart from tedium) in working with the outer horizon of non-extremal Reissner--Nordstr\"om or indeed any spherically symmetric non-extremal black hole. Extremal black holes would seem to require a more subtle analysis. In a different direction, there are certainly purely technical issues arising in dealing with non-extremal Kerr and Kerr--Newman, a topic we hope to turn to in the future. 
We do not expect to encounter any fundamental issues with  non-extremal Kerr and Kerr--Newman, but the extremal case is again likely to be problematic.

%==============================================================

%=====================================================
\section*{Acknowledgements}
%=====================================================

RG was supported by a Victoria University of Wellington PhD Doctoral Scholarship.
\\
MV was directly supported by the Marsden Fund, 
via a grant administered by the Royal Society of New Zealand.

%\clearpage
%=====================================================
%===================================================== 

%==================================================================
\end{document}